\documentclass[aps,prb,twocolumn,reprint,superscriptaddress,citeautoscript]{revtex4-2}

\usepackage{physics}
\usepackage{graphicx}
\usepackage{amsmath,amssymb}
\usepackage[normalem]{ulem}
\usepackage{bm}

\usepackage{xcolor}

\usepackage{hyperref}% add hypertext capabilities
\hypersetup{
	citecolor = blue,
	colorlinks = true,
	urlcolor = blue
}

\begin{document}

\title{Curvature-induced valley-dependent spin-orbit interaction}
	
\author{Ai Yamakage}
\email[]{ai@st.phys.nagoya-u.ac.jp}
\affiliation{Department of Physics, Nagoya University, Nagoya 464-8602, Japan}
\author{T. Sato}
\email[]{sato-tetsuya163@g.ecc.u-tokyo.ac.jp}
\affiliation{Institute for Solid State Physics, University of Tokyo, Kashiwa 277-8581, Japan}
\author{R. Okuyama}
\affiliation{Department of Physics, Meiji University, Kawasaki 214-8571, Japan}
\author{T. Funato}
\affiliation{Center for Spintronics Research Network, Keio University, Yokohama 223-8522, Japan}
\affiliation{Kavli Institute for Theoretical Sciences, University of Chinese Academy of Sciences, Beijing, 100190, China.}
\author{W. Izumida}
\affiliation{Department of Physics, Tohoku University, Sendai 980-8578, Japan}
\author{K. Sato}
\affiliation{National Institute of Technology, Sendai College, Sendai 989-3128, Japan}
\author{T. Kato}
\affiliation{Institute for Solid State Physics, University of Tokyo, Kashiwa 277-8581, Japan}
\author{M. Matsuo}
\affiliation{Kavli Institute for Theoretical Sciences, University of Chinese Academy of Sciences, Beijing, 100190, China.}
\affiliation{CAS Center for Excellence in Topological Quantum Computation, University of Chinese Academy of Sciences, Beijing 100190, China}
\affiliation{Advanced Science Research Center, Japan Atomic Energy Agency, Tokai, 319-1195, Japan}
\affiliation{RIKEN Center for Emergent Matter Science (CEMS), Wako, Saitama 351-0198, Japan}
	
\begin{abstract}
We construct a general theoretical framework for describing curvature-induced spin-orbit interactions on the basis of group theory. 
Our theory can systematically determine the emergence of spin splitting in the band structure according to symmetry in the wavenumber space and the bending direction of the material. 
As illustrative examples, we derive the curvature-induced spin-orbit coupling for carbon and silicon nanotubes. Our theory offers a strategy for designing valley-contrasting spin-orbit coupled materials by tuning their curvatures. 
\end{abstract}
	
\maketitle

\section{Introduction} 

% Izumida @ 2024/09/27
The spin-orbit interaction (SOI) in solids lies at the heart of the ability to manipulate electron spins in spintronics. It has been utilized through the spin Hall effect, a charge-to-spin conversion phenomenon, to generate, manipulate, and detect spin currents by various electrical means~\cite{sinova2015spin}. It also enables control of the magnetization by transferring spin as a torque from materials such as heavy metals, antiferromagnets, as well as oxide, topological~\cite{manchon2015new,manchon2019current}, and chiral materials~\cite{yang2021chiral}. In addition, valleytronics, which utilizes the valley degree of freedom in materials, has stimulated researchers' interest because the interplay between spin and valley plays a crucial role in quantum transport phenomena~\cite{Vitale2018,Ciarrocchi2022,sierra2021van}.
	
While the design of materials for a suitable SOI is an important subject in modern spintronics, the choices of elements and crystal structures are limited. As another strategy, researchers have extensively studied tuning the strength of the SOI by using the mechanical properties of the materials, such as strain. Strain-induced SOI has been studied in semiconductor quantum wells with static~\cite{bernevig2005spin,hruska2006Effectsa,li2008Strainassisted,norman2010mapping,liu2011Spin,matsuda2011Effect,norman2014CurrentInduced} and dynamical~\cite{sogawa2001Transport,sanada2011acoustically} strains. However, the application of strain engineering has been restricted to inherently strong SOI materials, because strain does not directly couple to spins. To overcome the previous limitations, we propose an alternate route for designing valley-dependent SOI by using the curvature of the material.
	
\begin{figure}
\centering
\includegraphics[scale=0.4]{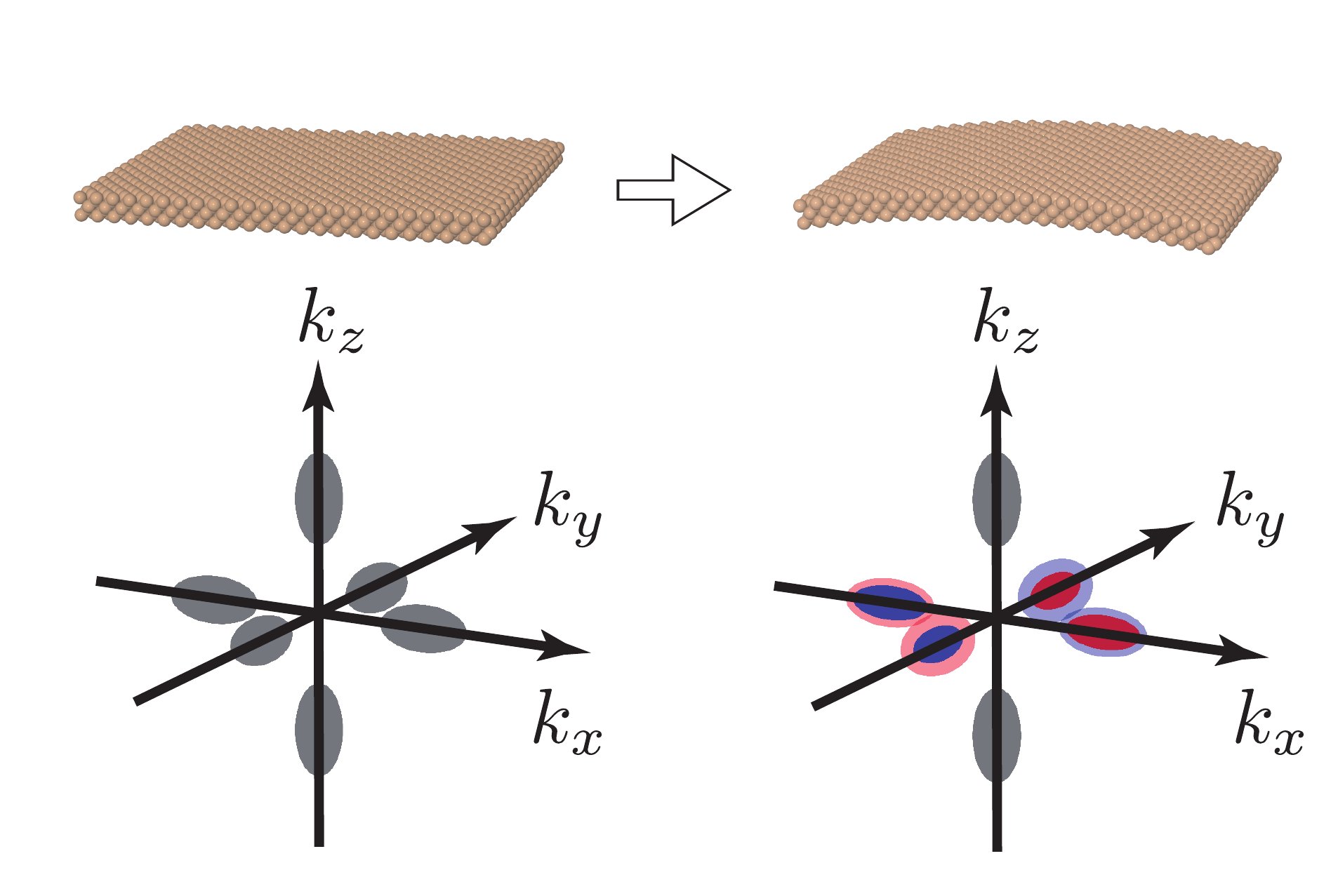}
\caption{Schematic diagram of valley-dependent spin splitting induced by the curvature of the material.
As shown in the left panel, we consider a system where Fermi surfaces exist along the $k_x$, $k_y$, and $k_z$ axes. 
In cubic crystals, all Fermi surfaces are equivalent; however, when curvature is introduced into the system, spin-split Fermi surfaces that depend on the valley emerge, as shown in the right panel.
}
\label{fig:schematic}
\end{figure}
	
In particular, we propose a novel strategy to realize valley-dependent SOI for electrons by breaking the crystalline symmetry through the application of a finite curvature to a material as shown in Fig.~\ref{fig:schematic}. Our theoretical framework provides a powerful method for determining the emergence of spin splitting in the band structure depending on high-symmetry points in the wavenumber space and the bending direction of the material. Our procedure based on group theory generally describes valley-dependent spin splitting induced by the curvature. We consider a single-wall carbon nanotubes (CNTs) and silicon as prominent examples. For CNTs, we derive an effective spin- and valley-dependent Hamiltonian, with an SOI allowed by the group theoretic approach, that is consistent with microscopic theory~\cite{Izumida2009-gx}. For silicon, we derive an effective model for conduction electrons near six valleys that clarifies the valley-dependent SOI induced by the curvature. 
Our work provides a general strategy for inducing SOI in materials and designing valley contrasting spintronics via tuning of the bending directions of material.

{The rest of this work is organized as follows. 
First, we briefly summarize a general procedure based on the group theory in Sec.~\ref{sec:GeneralProcedure} and 
outline a simplified tight-binding method for materials with a finite curvature in Sec.~\ref{sec:TightBindingCalculation}.
Next, we discuss the curvature-induced spin orbit interactions in CNTs and silicon in Sec.~\ref{sec:CNT} and Sec.~\ref{sec:Si}, respectively.
We also point out that a method of thin-film quantization employed in previous studies assumes unphysically large confinement potential in Sec.~\ref{sec:ThinFilmQuantization}.
Finally, we summarize our study in Sec.~\ref{sec:Summary}.}

\section{Group Theory} 
\label{sec:GeneralProcedure}

{The effects of spin-orbit interaction linked with curvature control should be approached in the most robust and universal manner, as opposed to an ad hoc methodology --- this is precisely the group theoretical method we have developed in this study.
The group-theoretic methods have played a vital role in discussions of spin-orbit interaction due to the symmetry of crystals and their deformation by strain, especially in semiconductor physics~\cite{winkler2003spin}. We expand this group-theoretic method to include the effects of curvature, represented by spatial {\it second} derivatives, and show that the curvature directly couples with spin degrees of freedom.
For this purpose, we extend the group-theoretic approach by including the curvature $\partial_i \partial_j u_k$ ($i,j,k=x,y,z$) in its procedure, where $u_k$ is a lattice displacement.
We itemize our group-theoretic procedure to generate an effective Hamiltonian induced by curvature that can be applied to arbitrary systems including ones with multiple internal degrees of freedom (spin, orbital, and sublattice) as follows:}
\begin{enumerate}
\item Find the space group, $k$ points, and irreducible representations (irreps) of the low-lying excitations in the flat (uncurved) system.
\item Set the irreps of the symmetry operations.
\item Compute the irreducible decomposition of the matrices (operators) $O_i$.
\item Compute the irreducible decomposition of the momentum $k_i$, strains $\partial_i u_j$ and curvatures $\partial_i \partial_j u_k$.
\item Construct the products of the matrix, momentum, strains, and curvatures. Decompose them into irreps. The totally symmetric representations can appear in the effective Hamiltonian. The other irreps correspond to physical observables such as electric/spin currents. 
\end{enumerate}
Our approach broadens the spectrum of materials, beyond strong spin-orbit materials utilized for spintronic devices, laying the foundation for a new era of `curvature engineering' and would greatly facilitate the study of curvature-induced effects across a variety of materials~{\footnote{{The curvature effect described by the second derivative of the displacement is analogious to the spin-vorticity coupling~\cite{matsuo2013Mechanical,kobayashi2017Spin,tateno2020Electrical}, $H_{\rm sv} = -\boldsymbol{s} \cdot \boldsymbol{\omega}/2$, where $\omega_{i} = \epsilon_{ijk} \partial_j \partial_t u_k$ means vorticity, dynamical antisymmetric lattice distortion. The vorticity can be regarded as a kind of curvature that extends the second-order derivative of the lattice displacements in space to those in space-time. In this sense, our group-theoretic approach can be extended to include spin-vorticity theory in a unified way.}}}.

\section{Tight-binding method}
\label{sec:TightBindingCalculation}

{A tight-binding calculation is a powerful method to clarify the curvature-induced SOI~\cite{Izumida2009-gx}.
In this section, we introduce a simplified tight-binding method to numerically estimate the curvature-induced spin splitting for curved thin films.}

\begin{figure}[tb]
\centering
\includegraphics[width=\linewidth]{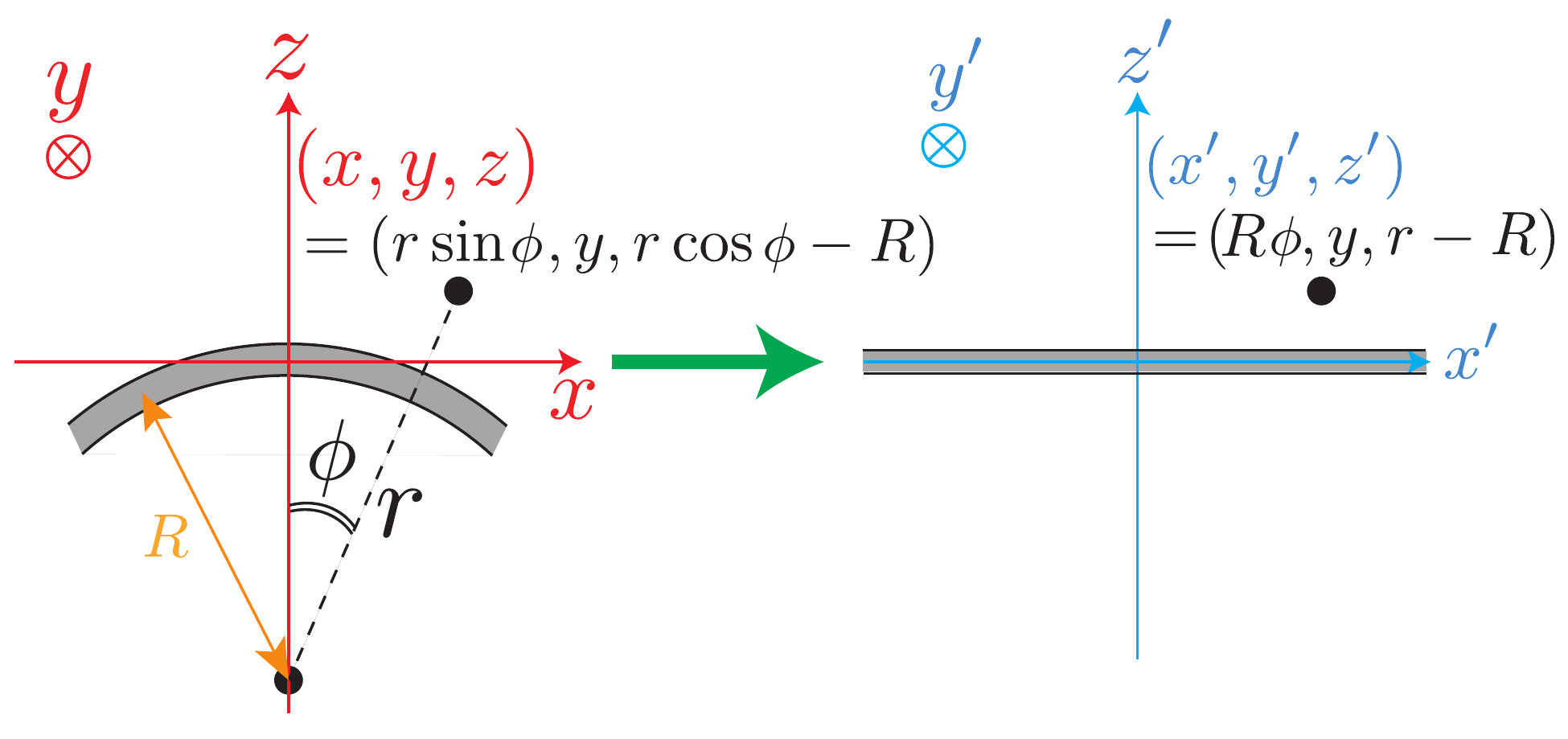}
\caption{Schematic diagram of the coordinate transformation. 
The gray region indicates a thin material.
Through this coordinate transformation, the curved material in the laboratory coordinates $(x,y,z)$ is mapped to a flat one in the new coordinates $(x',y',z')$.}
\label{fig:transformation}
\end{figure}

{Let us consider a curved thin material with curvature $1/R$, which is indicated by the gray area in the left panel of Fig.~\ref{fig:transformation}.
The curvature radius $R$ is assumed to be much larger than the thickness of the material.
Then, we introduce a general coordinate transformation to make a curved thin film flat in new coordinates (see  Fig.~\ref{fig:transformation}).}
The Cartesian coordinates at an arbitrary position are expressed in the cylindrical coordinates as $(x,y,z)=(r\sin\phi,y, r\cos\phi-R)$, where $r$ is the distance from the center of the curved material and $\phi$ is the azimuth angle measured from the $z$-axis (see Fig.~\ref{fig:transformation}).
We consider a general transformation from the original Cartesian coordinates $(x,y,z)$ to curved coordinates $(x',y',z')$, in which the curved thin material is mapped to a flat thin material, keeping the $y$ coordinate unchanged ($y=y'$).
The new coordinates are written with the cylindrical coordinates as $(x',y',z')=(R\phi, y,r-R)$, which leads to the relation,
\begin{align}
\begin{cases}
x=(z'+R)\sin (x'/R) , \\ y=y',\\ z=(z'+R)\cos (x'/R)-R .
\end{cases}
\label{eq:trans}
\end{align}
{In the new coordinates, the crystal structure has no curvature and its electronic structure can be obtained using the standard tight-binding calculation.
As compensation, both of the hopping energy between neighboring sites and the atomic SOI are modified due to the curvature of the coordinates.
We formulate these corrections up to $\mathcal{O}(1/R)$ (for a detail, see Appendix~\ref{app:calc}) and calculate the band structure by diagonalizing the tight-binding Hamiltonian with modified parameters of the hopping and SOI.
The results of the tight-binding model for CNTs and silicon tubes are given in Sec.~\ref{sec:calc-carbon} and Sec.~\ref{sec:calc-silicon}, respectively.}

\section{Carbon nanotube}
\label{sec:CNT}

The investigation of curvature-induced spin-orbit interaction in carbon nanotubes has been thoroughly explored through numerical calculations~\cite{Chico-2004-10,Chico-2009-06,Izumida2009-gx}.
To understand its physical origin, effective Hamiltonians derived from the perturbation theory have been utilized in several theoretical works~\cite{Ando-2000-06,Huertas-Hernando-2006-10,Izumida2009-gx,Jeong-2009-08,Ochoa2012}.
However, this approach can be a demanding task to include all the terms in the perturbation calculation.
Early theories indeed fell short in this regard, overlooking a key effect~\cite{Ando-2000-06,Huertas-Hernando-2006-10} that was later observed in experimental studies~\cite{Kuemmeth-2008-03} and subsequently identified in subsequent research~\cite{Izumida2009-gx,Jeong-2009-08}.
These challenges clarify necessity for developing a universal methodology for obtaining effective models that account for curvature effects.
To address this demand, we introduce a universal and robust method in this study, employing group theory in the construction of the Hamiltonian {in Sec.~\ref{sec:CNTGroupTheory} and \ref{subsub:Hamiltonian}.
Furthermore, we perform the tight-binding calculation to estimate the strength of the SOI induced by the curvature in Sec.~\ref{sec:calc-carbon}.}

\subsection{Group theory}
\label{sec:CNTGroupTheory}

\subsubsection{Irrep of low-energy states}

For nonmagnetic graphene, which is the flat (unrolled) system of CNT, characterized by the magnetic space group $P6/mmm1'$ (No.~191.234), the low-lying excitations are governed by the Dirac cones around the $K$ and $K'$ points, on which the magnetic little cogroup is given by $6'/mmm'$ and its unitary maximum subgroup $\bar 62m$ ($D_{3h}$)~\cite{BilbaoCrystallographicServer1,BilbaoCrystallographicServer2}.
They come from the $p_z$ orbitals to form a double degeneracy, single-valued two-dimensional odd-parity irrep $K_6$. 

\subsubsection{Symmetry operation}

The $K_6$ irreps of the generators, i.e., the $3^+$ threefold rotation along the $\ev{001}$ ($z$) direction, $2_{100}$ twofold rotation along the $\ev{100}$ ($y$) direction, and $m_{001}$ horizontal mirror, are listed as
follows~\cite{BilbaoCrystallographicServer1,BilbaoCrystallographicServer2,Xu2020-lo, Elcoro2021-gs}:
\begin{align}
D_K(3^+) &= e^{i \sigma_z 2\pi/3}, \\
D_K(2_{100}) &= \sigma_y, \\
D_K(m_{001}) &= -\sigma_0,
\end{align}
where $\sigma_0$ and $\sigma_i$ ($i=x,y,z$) denote the identity and Pauli matrices acting in the sublattice space, respectively.
Additionally, we have the parity-time (PT) symmetry operation,
\begin{align}
O \to \sigma_x O^* \sigma_x.
\end{align}
Taking the spin degrees of freedom into account, the irreps are given by
\begin{align}
\bar D_K(3^+) &= e^{i \sigma_z 2\pi/3} e^{-i s_z \pi/3}, \\
\bar D_K(2_{100}) &= \sigma_y (-i s_y), \\
\bar D_K(m_{001}) &= -\sigma_0 i s_z,
\end{align}
and {the PT symmetry operation} by
\begin{align}
O \to \sigma_x s_y O^* s_y \sigma_x.
\end{align}

\subsubsection{Irreps of matrices}

The theory on the $K$ point has the sublattice $\boldsymbol{\sigma}$ and spin $\boldsymbol{s}$ degrees of freedom, represented by $4 \times 4$ matrix $\sigma_\mu s_\nu$. 
This matrix transforms as $O \to D^\dag O D$. 
They are decomposed into irreps under the point group $\bar 62m$ ($D_{3h}$). 
Furthermore, they are decomposed into PT--even and odd irreps. 
The results are summarized in Table~\ref{d3h_1} {and \ref{curvature}. We note that the decompositions of the momentum and spatial derivative of the lattice displacement are also shown in these tables (see Sec.~\ref{IrrepsMSC}).}

Note that we will use the following convention for the components of the two-dimensional irreps, $E'$ and $E''$. Their components, $(E(1), E(2))$, are transformed for $g \in D_{3h}$ as if they are components of in-plane geometric vectors:
\begin{align}
\bar D_K(3^+)^\dag 
\begin{pmatrix} E(1) \\ E(2) \end{pmatrix}
\bar D_K(3^+) &= \hat{R}(2\pi/3) \begin{pmatrix} E(1) \\ E(2) \end{pmatrix}, \\
\bar D_K(2_{100})^\dag
\begin{pmatrix} E(1) \\ E(2) \end{pmatrix} 
\bar D_K(2_{100}) &= 
\begin{pmatrix} -E(1) \\ E(2) \end{pmatrix},
\end{align}
where $\hat{R}(\vartheta)$ is a two-dimensional rotation matrix,
\begin{align}
\hat{R} (\vartheta) = 
\begin{pmatrix}
\cos \vartheta & - \sin \vartheta \\
\sin \vartheta & \cos \vartheta
\end{pmatrix}.
\end{align}
By using this convention, the product rules involving $E'$ and $E''$ irreps are simplified as shown in Table \ref{d3h_prod}. These results can be easily interpreted geometrically: $A'_1$ and $A'_2$ components of $E' \times E'$ are inner and outer products of the two in-plane vectors, respectively.

\begin{table*}[tb]
\caption{Irreps of matrices under $\bar 6 2 m$ ($D_{3h}$) point group. Irreps of momenta and strains are also shown.}
\begin{ruledtabular}
\begin{tabular}{lrrrrccccc}
Irrep & $1$ & $3^+$ & $2_{100}$ & $m_{001}$ & PT--even & PT--odd & $\vb*{k}$ & $\vb*{kk}$ & $\grad{\vb*{u}}$ \\
\hline
$A_1'$ & 1 & 1 & 1 & 1 & 1, $\sigma_z s_z$ & & & $k_x^2+k_y^2$ & $\partial_x u_x + \partial_y u_y$ \\
$A_2'$ & 1 & 1 & $-1$ & 1 & & $\sigma_z$, $s_z$ & & & $\partial_x u_y - \partial_y u_x$ \\
$E'$ & 2 & $-1$ & 0 & 2 & $(\sigma_x, \sigma_y)$ & $(\sigma_y, -\sigma_x) s_z$ & $(k_x, k_y)$ & $(2k_xk_y, k_x^2-k_y^2)$ & $(\partial_x u_y + \partial_y u_x, \partial_x u_x - \partial_y u_y)$ \\
$A_1''$ & 1 & 1 & 1 & $-1$ & & $\sigma_x s_x + \sigma_y s_y$ \\
$A_2''$ & 1 & 1 & $-1$ & $-1$ & & $\sigma_x s_y - \sigma_y s_x$ & \\
$E''$ & 2 & $-1$ & 0 & $-2$ & $\sigma_z (s_y, -s_x)$ & $(s_x, s_y)$,  & & & $(\partial_y, -\partial_x)u_z$ \\
& & & & & & $(\sigma_x s_y + \sigma_y s_x, \sigma_x s_x - \sigma_y s_y)$ 
\end{tabular}
\end{ruledtabular}
\label{d3h_1}
\end{table*}

\begin{table}
\caption{Irreps of curvatures.}
\begin{ruledtabular}
\begin{tabular}{lc}
Irrep & $\grad\grad{\vb*{u}}$ \\
\hline
$A_1'$ & $(\partial_x^2-\partial_y^2)u_y + 2 \partial_x \partial_y u_x$ \\
$A_2'$ & $(\partial_x^2 - \partial_y^2) u_x - 2\partial_x \partial_y u_y$ \\
$E'$ & $\laplacian (u_x, u_y)$, \\
& $\qty(2\partial_x \partial_y u_y + (\partial_x^2-\partial_y^2)u_x, 2\partial_x\partial_y u_x - (\partial_x^2-\partial_y^2)u_y)$ \\
$A_1''$ & \\
$A_2''$ & $\laplacian{u_z}$ \\ 
$E''$ & $(\partial_y^2-\partial_x^2, 2\partial_x\partial_y)u_z$
\end{tabular}
\end{ruledtabular}
\label{curvature}
\end{table}

\begin{table}
\caption{Product rules for two-dimensional irreps of $D_{3h}$. $X$ indicates $X'$ or $X''$ in the table ($X = A_1,A_2,$ or $E$). $X' \times Y' = X'' \times Y'' = Z'$ and $X' \times Y'' = X'' \times Y' = Z''$.}
\begin{tabular}{rl}
\hline \hline
\multicolumn{1}{l}{Product} & Components \\
\hline
$A_1 \times E=E\,\,\,$ & ${A_1} (E(1), E(2))$ \\
$A_2 \times E=E\,\,\,$ & ${A_2} (E(2), -E(1))$ \\
$E\,\, \times E=A_1$ & $E(1) E(1) + E(2) E(2)$ \\
$+A_2$ & $E(1) E(2) - E(2) E(1)$ \\
$+E\,\,$ & $(E(1) E(2) + E(2) E(1), E(1) E(1) - E(2) E(2))$ \\
\hline \hline
\end{tabular}
\label{d3h_prod}
\end{table}

\subsubsection{Irreps of momentum, strain, and curvature}
\label{IrrepsMSC}

A three-dimensional displacement $\vb*{u} = (u_x, u_y, u_z)$ is transformed using a symmetry operation $g$ of the little group as
\begin{align}
3^+ : \ \pmqty{ u_x \\ u_y \\ u_z } &\to 
\pmqty{-1/2 & -\sqrt{3}/2 & 0 \\
\sqrt{3}/2 & -1/2 & 0 \\
0 & 0 & 1}
\pmqty{ u_x \\ u_y \\ u_z }, \\
2_{100} : \ 
\pmqty{ u_x \\ u_y \\ u_z } &\to
\pmqty{ -1 & 0 & 0 \\
   0 & 1 & 0 \\
   0 & 0 & -1} 
\pmqty{ u_x \\ u_y \\ u_z }, \\
m_{001}: \ 
\pmqty{ u_x \\ u_y \\ u_z } &\to
\pmqty{ 1 & 0 & 0 \\
   0 & 1 & 0 \\
   0 & 0 & -1}
\pmqty{ u_x \\ u_y \\ u_z }.
\end{align}
Two-dimensional vectors, momentum $\vb*{k} = (k_x, k_y)$ and gradient $\grad = (\partial_x, \partial_y)$, transform in a similar way. 
Note that momentum is PT--even, while the gradient and displacement are PT--odd.
The irreducible decomposition of strains and linear and quadratic polynomials of momentum is shown in Table~\ref{d3h_1}, whereas that of curvatures is shown in Table~\ref{curvature}.

\subsection{Effective Hamiltonian}
\label{subsub:Hamiltonian}

Now we are in a position to obtain any observable operator, including the Hamiltonian. 
The Hamiltonian must belong to the totally symmetric irrep, $A_1'$. 
Of the order of $k^0$ and on the flat space, the Hamiltonian is given by a linear combination of $\sigma_0 s_0$ and $\sigma_z s_z$, the latter of which corresponds to spin-orbit coupling in graphene leading to a quantum spin Hall insulator~\cite{Kane2005-on} and is ignored in this work because the effect is quite small in the CNTs.
Since both $(k_x, k_y)$ and $(\sigma_x, \sigma_y)$ belong to the PT--even $E'$ irrep, the Hamiltonian of the first order of $k$ is constructed as the product of them as $\propto k_x \sigma_x + k_y \sigma_y$, that is
\begin{align}
H_K^{(0)}(\boldsymbol{k}) = \hbar v_{\rm F} (k_x \sigma_x + k_y \sigma_y),
\end{align}
which is nothing but a Dirac cone.

\subsubsection{Strain-induced chiral gauge field}

Strain $\partial_i u_j$ can appear in the Hamiltonian when the product of $\partial_i u_j$ and matrices are in the PT--even $A'_1$ irrep. 
The $A'_2$ strain, $\partial_x u_y - \partial_y u_x$, is not allowed in the Hamiltonian because there is no PT--even $A_2'$ matrix. 
The $E'$ strains, on the other hand, can appear as
\begin{align}
 \propto \qty(\partial_x u_y + \partial_y u_x) \sigma_x + \qty(\partial_x u_x - \partial_y u_y) \sigma_y,
\end{align}
reproducing the chiral gauge field in strained graphene~\cite{Vozmediano2010-je}. 
The $E''$ strain, $(\partial_y, -\partial_x) u_z$, also appears in the Hamiltonian to couple with PT--even spin-dependent $E''$ matrix $\sigma_z( s_y, -s_x)$:
\begin{align}
 \propto \partial_y u_z \sigma_z s_y + \partial_x u_z \sigma_z s_x,
\end{align}
which is the mass term for the Dirac cone with spin dependence. 
Further strain effect will not be discussed below since it is outside the scope of this paper.

\subsubsection{Curvature-induced spin-orbit interaction.}

The next-order terms, $\partial_i \partial_j u_k$, including curvature, are coupled with PT--odd matrices in the Hamiltonian, resulting in spin-split eigenstates. 
For instance, the $A_2''$ and $E''$ terms are included in the Hamiltonian in the form of
\begin{align}
\propto & \ \laplacian{u_z} \qty(\sigma_x s_y - \sigma_y s_x), \ 
\qty(\partial_y^2 - \partial_x^2) u_z s_x + 2\partial_x \partial_y u_z s_y, \nonumber \\
& \qty(\partial_y^2 - \partial_x^2) u_z 
\qty(\sigma_x s_y + \sigma_y s_x) + 2\partial_x \partial_y u_z \qty(\sigma_x s_x - \sigma_y s_y),
\end{align}
{which contributes} to the curvature-induced spin-orbit interaction as discussed below.

\begin{figure}
\centering
\includegraphics[scale=0.60]{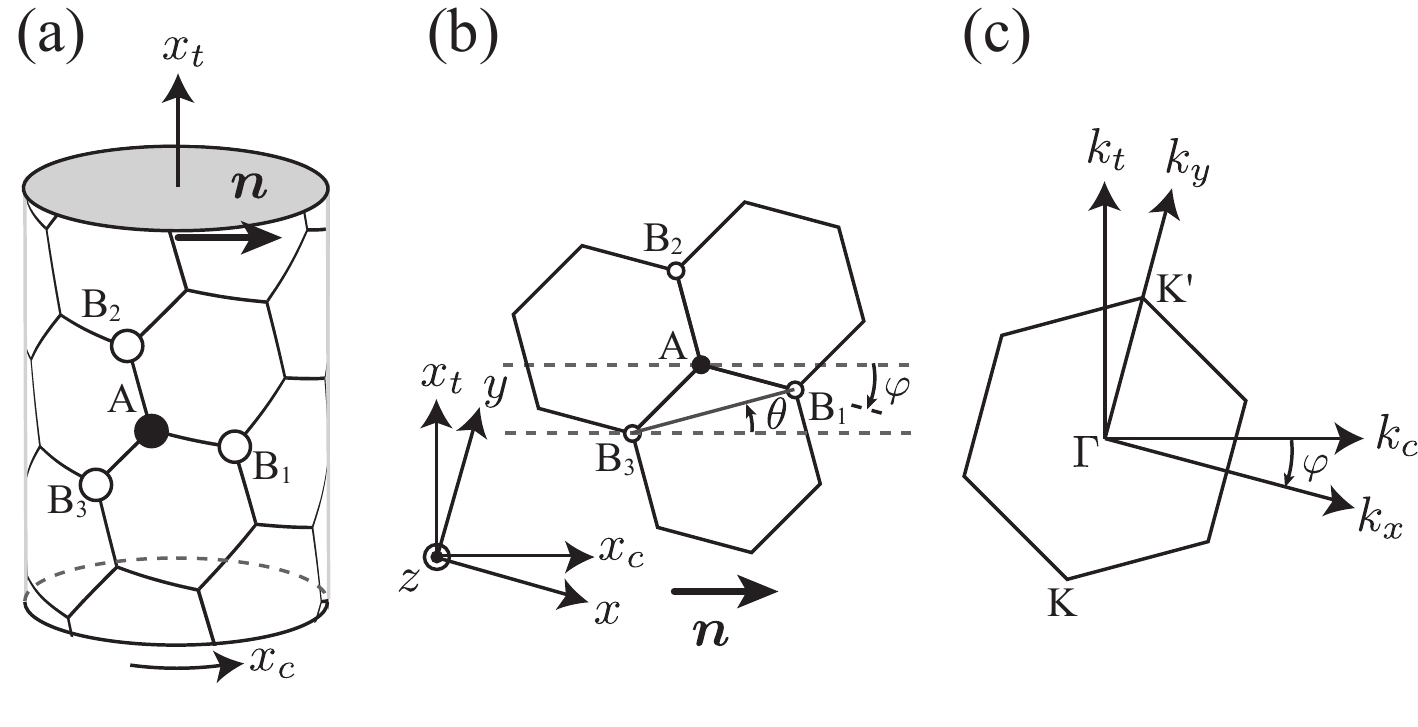}
\caption{(a) Atomic structure of a cylindrical surface. (b) Coordinates of CNT in real space. (c) Coordinates of CNT in momentum space.}
\label{Fig2}
\end{figure}

\begin{table*}[t]
\caption{Irreducible decomposition under $D_{3h}$ of momentum $\boldsymbol k$, spin $\boldsymbol{s}$, pseudospin for sublattice $\boldsymbol{\sigma}$, and $\partial_i \partial_j u_z$ of the order of $1/R$ (see Eq.~(\ref{u_z})). The twofold axis of $C_2'$ is set to $y$. 
The momentum $\boldsymbol{k}$ is PT-even and {$\partial_i \partial_j u_z$}
is PT-odd.}
\begin{tabular}{cllll}
\hline\hline
Irrep & PT-even & PT-odd & $1/R$ (PT-odd) & $1/R^2$ (PT-even) \\
\hline 
$A_1'$  & $\sigma_z s_z$ & & & $1/R^2$ \\
$A_2'$  & & $\sigma_z$, $s_z$ \\
$E'$ &  $(k_x, k_y)$, $(\sigma_x, \sigma_y)$ & $(\sigma_y s_x, -\sigma_x s_z)$ & & $(\sin2\varphi, \cos2\varphi)/R^2$, $(\sin4\varphi, -\cos4\varphi)/R^2$ \\
$A_1''$ & & $\sigma_x s_x + \sigma_y s_y$ \\
$A_2''$ & & $\sigma_x s_y - \sigma_y s_x$ & $1/R$ \\
$E''$ & $(\sigma_z s_y, -\sigma_z s_x)$ & $(s_x, s_y)$, & $(\cos2\varphi, -\sin2\varphi)/R$ \\
& & $(\sigma_x s_y + \sigma_y s_x, \sigma_x s_x - \sigma_y s_y)$ \\
\hline\hline
\end{tabular}
\label{d3h}
\end{table*}
	
The curvature $1/R$, where $R$ is the radius of the carbon nanotube, is expressed in terms of the displacement field $\boldsymbol{u}(\boldsymbol{x})$, where $\boldsymbol{x} = (x, y, 0)$ is the coordinates of the tube surface. For the tube geometry shown in Fig.~\ref{Fig2}, the displacement field is given by
\begin{align}
\boldsymbol{u}(\boldsymbol{x}) = \qty(0, 0, u_z(\boldsymbol{x} \cdot \boldsymbol{n})),
\end{align}
using the unit vector of the bending direction $\boldsymbol{n}= (\cos\varphi, \sin\varphi, 0)$, where $\varphi = \pi/6-\theta$ and $\theta$ is the chiral angle (see Fig.~\ref{Fig2}~(b)). The second derivative $\partial_i \partial_j u_z$ is proportional to the first order of the curvature:
\begin{align}
& (\partial_x^2 + \partial_y^2) u_z \propto \frac{1}{R}, \notag \\
& (\partial_x^2 - \partial_y^2) u_z \propto \frac{\cos2\varphi}{R}, \
2 \partial_x \partial_y u_z \propto \frac{\sin2\varphi}{R}.
\label{u_z}
\end{align}
We re-summarize the table of the irreducible decomposition with the first- and second-order curvatures in Table~\ref{d3h} for convenience of the present discussion.
The coupling terms in the Hamiltonian between the curvature and other quantities are obtained as PT-even irreps of $A_1'$: 
{\begin{align}
\propto \ & (\sigma_x s_y - \sigma_y s_x)/R, \ 
(s_x \cos2\varphi - s_y \sin2\varphi)/R, \notag \\
& [\cos2\varphi\, (\sigma_x s_y + \sigma_y s_x) - \sin2\varphi\, (\sigma_x s_x - \sigma_y s_y)]/R, \notag \\
& (\sigma_x \sin2\varphi + \sigma_y \cos2\varphi)/R^2, \notag \\
& (\sigma_x \sin4\varphi - \sigma_y \cos4\varphi)/R^2.
\end{align}}

To simplify the Hamiltonian, we introduce the tube coordinates $(k_c, k_t)^{\mathrm T} = \hat{R}(-\varphi) (k_x, k_y)^{\mathrm T}$, with $k_c$ and $k_t$ being momenta along the circumference and axis directions, respectively {(see Fig.~\ref{Fig2}(c))}.
$\boldsymbol{\sigma}$ and $\boldsymbol{s}$ are also represented in this basis as $(\sigma_c, \sigma_t)^{\mathrm T} = \hat{R}(-\varphi) (\sigma_x, \sigma_y)^{\mathrm{T}}$ and $(s_c, s_t)^{\mathrm T} = \hat{R}(-\varphi) (s_x, s_y)^{\mathrm{T}}$. 
Then, we can construct the effective Hamiltonian in the presence of the curvature-induced SOI as $H_K (\boldsymbol{k})$ by collecting the terms with PT-even irrep of $A_1'$. The effective Hamiltonian for the opposite valley, $K'$, is obtained by requiring the Hamiltonian to be even under time reversal: $T H_K(\boldsymbol{k}) T^{-1} = H_{K'}(-\boldsymbol{k})$. Here, we have used $T = i \tau_x \sigma_z s_y \mathcal{K}$, with $\boldsymbol{\tau}$ being the valley-pseudospin operator~\footnote{The time reversal operator $T$ for our basis (same as the one used in Ref.~\cite{Izumida2009-gx}) is derived from the well-known expression, $T = -i s_y \mathcal{K}$, for the position basis.}. Finally, we obtain the valley-dependent effective Hamiltonian as,
\begin{align}
& H_{\tau_z K}(\boldsymbol{k}) \notag \\
& = \hbar v_{\mathrm{F}} \qty(k_{c} \sigma_{c} + \tau_z k_t \sigma_t) -\epsilon_{\mathrm{so}}\tau_z \qty(s_c\cos3\varphi-s_t \sin3\varphi) \notag \\
& - \hbar v_{\mathrm{F}} \Delta k_{\mathrm{so}} (\sigma_c s_t - \tau_z \sigma_t s_c)
- \hbar v_{\mathrm{F}} \Delta k_{\mathrm{so}}' (\sigma_c s_t  + \tau_z \sigma_t s_c) \notag \\
& - \hbar v_{\mathrm{F}} \qty(\tau_z \sigma_c \Delta k_c \sin 3\varphi + \sigma_t \Delta k_t \cos3\varphi).
\end{align}
Here, $\tau_z K$ represents $K$ and $K'$ for $\tau_z = \pm1$. In the curvature-induced SOI terms, $\epsilon_{\mathrm{so}}$, $\Delta k_{\mathrm{so}}$, and $\Delta k'_{\mathrm{so}}$ are proportional to $1/R$. 
The spin-independent shift ($\Delta k_{c}$, $\Delta k_t$) of Dirac points is proportional to $1/R^2$. 
The obtained Hamiltonian is consistent with the previous study~\footnote{For the perturbative calculation given in Ref.~\cite{Izumida2009-gx}, we have $\Delta k'_{\rm SO} = 0$. 
This is due to the oversimplification in the model, i.e., the artificial electron-hole symmetry for $\pi$ and $\sigma$ orbitals. $\Delta k'_{\rm SO} \neq 0$ is obtained from more realistic analyses.}. We should note that the chiral-angle dependence of the CNT is also fully reproduced. This means that the $3\varphi$ dependence is ascribable to crystalline symmetry. In conclusion, the curvature breaks the spatial-inversion symmetry of graphene and induces the antisymmetric SOI which leads to valley-dependent spin-split energy bands.

\subsubsection{Valley-asymmetric velocities}

Similarly, we can reproduce the curvature-induced valley-asymmetric Fermi velocities of electrons that are microscopically derived in Ref.~\cite{Izumida2012}.
These are determined by terms proportional to $\boldsymbol{k}$ in the effective Hamiltonian.
For the $K$ valley, we decompose the products of sublattice pseudospin, momentum, and derivatives of $\boldsymbol{u}$ by neglecting tiny spin dependence.
We obtain the effective Hamiltonian as the PT--even irreps of $A'_1$, where the leading order in curvature is $1/R^2$.
By requiring the time-reversal symmetry, the additional terms for the Hamiltonian for $\tau_z K$ valley yield,
\begin{align}
H'_{\tau_z K} (\boldsymbol{k}) &= \frac{1}{R^2} \Bigl[ g_1 \sigma_c k_c + g_2 \tau_z \sigma_t k_t \notag \\
& \quad + g_3 \tau_z k_c \sin 3\varphi + g_4 \tau_z k_t \cos 3\varphi \notag \\
& \quad + g_5 (\sigma_c, - \tau_z \sigma_t) \hat{R}(-6\varphi)
\begin{pmatrix} k_c \\ k_t \end{pmatrix}
\Bigr],
\label{eq:H_av}
\end{align}
where $g_j$'s are constants independent of both $R$ and $\varphi$. 
Equation~\eqref{eq:H_av} is nothing but the Hamiltonian derived in Ref.~\cite{Izumida2012}.
Note that some coupling constants, which are independent in Ref.~\cite{Izumida2012}, satisfy the relation $-c_3 = c_4 = 4 g_5 / \hbar$ in our analysis.
This should be because the symmetries of the system may not be fully taken into account in the previous calculation based on perturbation theory.
Actually, $c_3$ and $c_4$ obtained from a fitting with the numerical calculations satisfy $-c_3 \simeq c_4$ in Ref.~\cite{Izumida2012}.

\subsection{Tight-binding calculation}
\label{sec:calc-carbon}

\begin{figure*}[htbp]
\centering
\includegraphics[width=150mm]{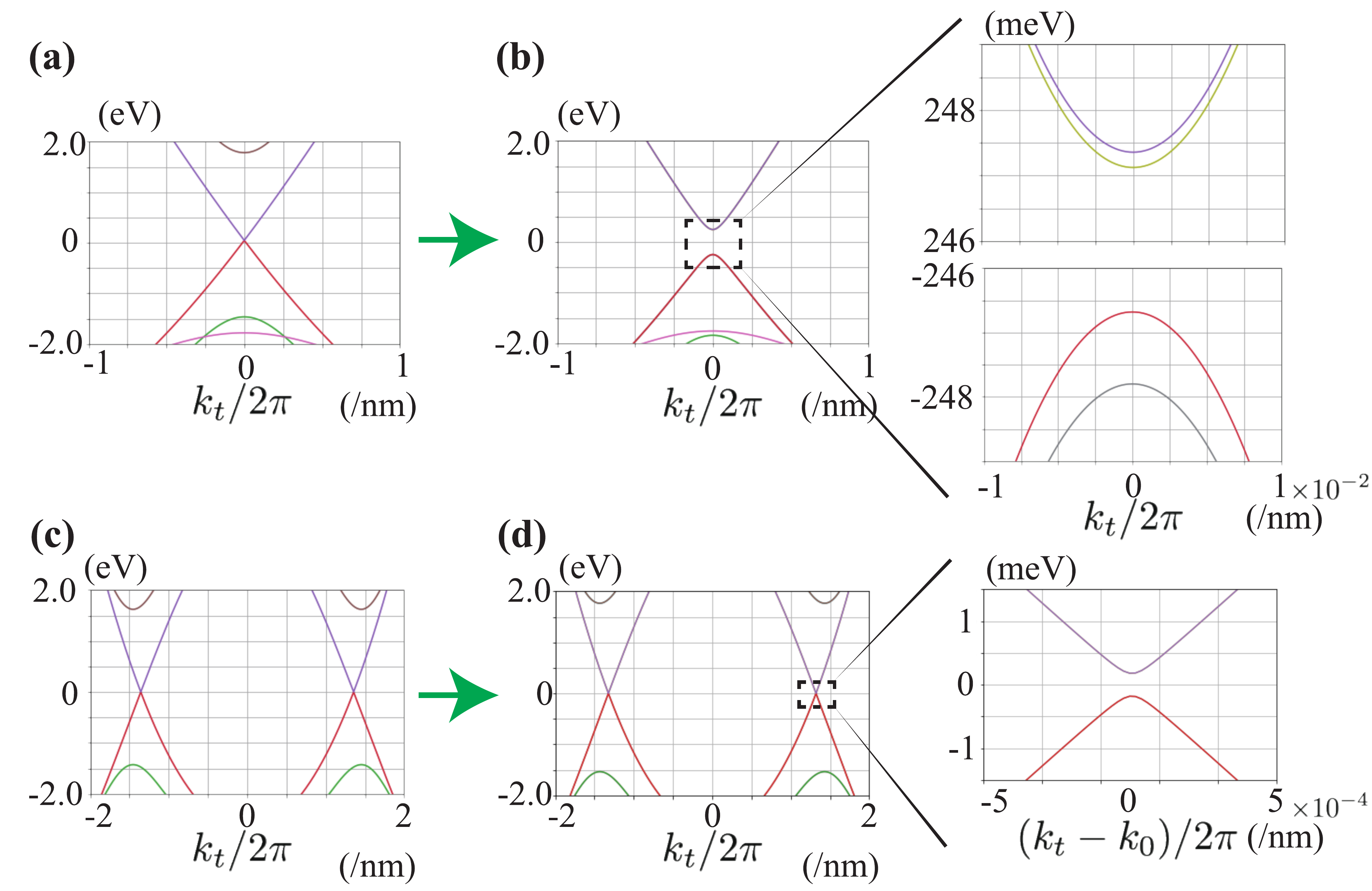}
\caption{Band structures for (a-b) $(9,0)$ zigzag nanotube and (c-d) $(6,6)$ armchair nanotube.
$k_0\simeq1.3225\,{\rm nm}^{-1}$ and $k_t$ is a momentum along the axis direction.
(a)(c) Band structures in the absence of curvature, which are obtained from that of graphene on the cutting lines, reflecting the boundary condition in the circumferential direction.
(b)(d) Band structures in the presence of curvature. A band gap opens up at the Dirac point in both the zigzag and armchair nanotubes.
The gap for the zigzag nanotube ($\sim 500\,{\rm meV}$) is much larger than that for the armchair ($\sim 0.4\,{\rm meV}$).
For the zigzag nanotube, spin splitting is observed in both the conduction band $(\sim 0.2\,{\rm meV})$ and the valence band ($\sim 1.1\,{\rm meV}$).
On the other hand, for the armchair nanotube, spin splitting is negligibly small in both bands. 
These results are consistent with Ref.~\cite{Izumida2009-gx}.}
\label{fig:carbon}
\end{figure*}

From our tight-binding calculation, we obtained the band structures for the low-lying cutting lines in the absence of curvature for $(9,0)$ (zigzag) and $(6,6)$ (armchair) nanotubes in Fig.~\ref{fig:carbon}~(a) and (c), respectively.
When the curvature is non-zero, the band structures of the zigzag and armchair nanotubes change into those in Fig.~\ref{fig:carbon}~(b) and (d), respectively.
Here, the radii of the $(9,0)$ and $(6,6)$ nanotubes are $R\simeq 0.35\,{\rm nm}$ and $R\simeq 0.41\,{\rm nm}$, respectively.
The band gap between the conduction and valence bands is estimated to be $500\,{\rm meV}$ for the $(9,0)$ nanotube and $0.4\,{\rm meV}$ for the $(6,6)$ nanotube.
The spin splitting energies for the conduction and valence bands are estimated to be $0.2\,{\rm meV}$ and $1.1\,{\rm meV}$ for the $(9,0)$ nanotube, while they become much smaller for the $(6,6)$ nanotube.
The band gap opens exactly at the $K$ point for $(9,0)$ nanotube, while its position is shifted from the $K$ point by $k_tT/2\pi \sim 0.08$ for $(6,6)$ nanotube, where $T=2\sqrt{3}\pi R/d_R$ is the length of the unit cell of the nanotube, and $d_R$ is the greatest common divisor of $2n+m$ and $2m+n$ for the $(n,m)$ nanotube~\cite{Saito-1998}.

These results semi-quantitatively agree with the previous theoretical work.
In fact, the band gap is opened by the curvature in the same way as in Ref.~\cite{Izumida2009-gx}.
The spin splitting of the $(9,0)$ nanotube and the gap of the $(6,6)$ nanotube, which are proportional to $1/R$, have similar values to those in Ref.~\cite{Izumida2009-gx}.
However, the gap of the $(9,0)$ nanotube, which is proportional to $1/R^2$, is several times larger than the value reported in Ref.~\cite{Izumida2009-gx}. 
This deviation is consistent with the fact that our calculation neglected the contribution of $1/R^2$; the present tight-binding calculation leads to a reasonable estimate only up to $1/R$.

\section{Silicon}
\label{sec:Si}

{Following the success of our theory for CNTs, we next} consider curvature-induced SOI in silicon, which is a fundamental material used in semiconductor technology. We focus on the lowest conduction bands, whose edges are located at $\Delta$ points near the zone boundary along the $\Gamma$-X symmetry lines, leading to valley structures in electron doped silicon.
In this section, we discuss the expected effect of curvature on, e.g., silicon nanotubes~\cite{Huang2013,Taghinejad2013}.

\subsection{Group theory}
	
Three-dimensional silicon crystallizes into a diamond structure whose symmetry is characterized by the space group $Fd\bar 3m$ (No.~227). The conduction electrons are located at the six valleys at $\Delta$ points, $(\pm k_0, 0, 0), (0, \pm k_0, 0)$, and $(0,0,\pm k_0)$ (see also Fig.~\ref{fig:schematic}). Their magnetic little group is $4/m'mm$ ($C_{4v} \times \{I, PT\}$). The orbital wavefunctions for the conduction band minima belong to $\Delta_1$ irrep~\cite{BilbaoCrystallographicServer2,footnote}. For $(\pm k_0, 0, 0)$-valley, the momentum $\boldsymbol k$ (measured from the band bottom), spin $\boldsymbol s$, and curvature $\partial_i \partial_j u_k$ decompose into the irreps summarized in Table~\ref{silicon_table}. 
The irreducible decomposition for the other valleys can be obtained by the threefold rotation, $(x,y,z) \rightarrow (y,z,x) \rightarrow (z,x,y)$, as shown in Table~\ref{tab:silicon_curvature}.
	
\begin{table}
\caption{Irreducible decomposition of momentum $\boldsymbol{k}$, spin $\boldsymbol{s}$, and $\partial_i \partial_j u_k$ of the order of $1/R$ under the point group $C_{4v}$, which is the little group of $(\pm k_0, 0, 0)$ from the space group $Fd \bar 3m$.}
\begin{ruledtabular}
\begin{tabular}{cllllll}
Irrep & PT-even & PT-odd & $\boldsymbol{\nabla}\boldsymbol{\nabla}\boldsymbol{u}$ (PT-odd) \\
\hline
$A_1$ & $k_x$ & & $\qty(\partial_y^2+\partial_z^2)u_x$, $\partial_x^2 u_x$, $\partial_x(\partial_y u_y + \partial_z u_z)$ \\
$A_2$ & & $s_x$ & $\partial_x \qty(\partial_y u_z - \partial_z u_y)$ \\
$B_1$ & & & $\qty(\partial_y^2-\partial_z^2)u_x$, $\partial_x (\partial_y u_y -\partial_z u_z)$ \\
$B_2$ & & & $\partial_y \partial_z u_x$, $\partial_x (\partial_y u_z + \partial_z u_y)$ \\
$E$ & $(k_y,k_z)$ & $(s_z, -s_y)$ & $\partial_x^2(u_y, u_z)$, \\ 
& & & $\partial_y \partial_z (u_y, u_z)$, $\qty(\partial_y, \partial_z)\partial_x u_x$, \\ 
& & & $\qty(\partial_y^2 u_y, \partial_z^2 u_z)$, $\qty(\partial_z^2 u_y, \partial_y^2 u_z)$
\end{tabular}
\end{ruledtabular}
\label{silicon_table}
\end{table}

\begin{table*}
\caption{Irreducible decomposition of momentum $\boldsymbol{k}$, spin $\boldsymbol{s}$, and curvature of order $1/R$, under the point group $C_{4v}$, which is the little group of six conduction valley minima, $(0,0,\pm k_0)$, $(\pm k_0, 0, 0)$, and $(0,\pm k_0, 0)$, from the space group $Fd \bar 3m$. We assume that curvature is expressed as $\boldsymbol{u} = (0,0,u_z(\boldsymbol{x} \cdot \boldsymbol{n}))$ with $\boldsymbol{n}= (\cos\varphi,\sin\varphi,0)$.}
\label{tab:silicon_curvature}
\begin{ruledtabular}
\begin{tabular}{c|ccc|ccc|ccc}
& \multicolumn{3}{c|}{$(0,0,\pm k_0)$} & \multicolumn{3}{c|}{$(\pm k_0,0,0)$} & \multicolumn{3}{c}{$(0,\pm k_0,0)$} \\
& $\boldsymbol{s}$ & $\boldsymbol{k}$ & $1/R$ &
$\boldsymbol{s}$ & $\boldsymbol{k}$ & $1/R$ &
$\boldsymbol{s}$ & $\boldsymbol{k}$ & $1/R$ \\
Irrep & (PT-odd) & (PT-even) & (PT-odd) & & & \\
\hline
$A_1$ & & $k_z$ & $1/R$ & & $k_x$ & & & $k_y$ & \\
$A_2$ & $s_z$ & & & $s_x$ & & $\sin(2\varphi)/R$ & $s_y$ & & $-\sin(2\varphi)/R$ \\
$B_1$ & & & $\cos(2\varphi)/R$ & & & & & & \\
$B_2$ & & & $\sin(2\varphi)/R$ & & & $\sin(2\varphi)/R$ & & & $\sin(2\varphi)/R$ \\
$E$ & $(s_y, -s_x)$ & $(k_x,k_y)$ & & $(s_z,-s_y)$ & $(k_y,k_z)$ & $(0,\cos^2\varphi/R)$, & $(s_x,-s_z)$ & $(k_z,k_x)$ & $(\sin^2\varphi/R,0)$, \\
& & & & & & $(0,\sin^2\varphi/R)$ & & & $(\cos^2\varphi/R,0)$
\end{tabular}
\end{ruledtabular}
\end{table*}
 
An effective Hamiltonian for the valley located at $\boldsymbol{k}_0$ is a totally symmetric irrep of the little group (PT-even irrep of $A_1$). In the absence of curvature, we obtain quadratic kinetic terms with anisotropic effective masses,
\begin{align}
H^{(0)}_{\boldsymbol{k}_0}(\boldsymbol{k}) &= 
\frac{\hbar^2 k_\ell^2}{2m_\ell}
+ \frac{\hbar^2 \boldsymbol{k}_t^2}{2m_t},
\end{align}
where $k_\ell$ and $\boldsymbol{k}_t$ are longitudinal (parallel to $\boldsymbol{k}_0$) and transversal (perpendicular to $\boldsymbol{k}_0$) momenta, respectively.
	
Hereafter, we consider curved silicon expressed with the displacement vector $\boldsymbol{u} = (0,0,u_z(\boldsymbol{x} \cdot \boldsymbol{n}))$ with $\boldsymbol{n}= (\cos\varphi,\sin\varphi,0)$, for simplicity. The curvature is represented by the second-order derivative $u_z''(\boldsymbol{x} \cdot \boldsymbol{n}) \propto 1/R$ as in the discussion on the CNT (see Eq.~\eqref{u_z}).
For $(\pm k_0, 0, 0)$-valley, as shown in Table~\ref{silicon_table}, we have a non-vanishing term of order $1/R$, $\partial_x (\partial_y u_z - \partial_z u_y) \propto \sin2\varphi/R $, as an $A_2$ irrep. 
This can be coupled to the same irrep, $s_x$, to be a totally symmetric irrep. Similarly, an $E$ irrep $(s_z, -s_y)$ can be coupled to $(\partial_x^2 u_y, \partial_x^2 u_z) \propto (0, \cos^2\varphi)/R$ and $(\partial_z^2 u_y, \partial_y^2 u_z) \propto (0, \sin^2\varphi)/R$. Accordingly, the curvature-induced SOI yields
\begin{align}
\label{eq:silicon}
& H'_{(\pm k_0, 0, 0)} \nonumber \\
&= \pm \frac{\hbar v_1}{R} s_x \sin 2\varphi 
\pm \frac{\hbar v_2}{R} s_y \cos^2\varphi 
\pm \frac{\hbar v_3}{R} s_y \sin^2\varphi.
\end{align}
Here, the double-sign is in the same order, and the signs are determined to satisfy time-reversal symmetry, $s_y H_{(k_0,0,0)}(\boldsymbol{k})^* s_y = H_{(-k_0,0,0)}(-\boldsymbol{k})$. 
For $(0, \pm k_0, 0)$-valley, the irreducible decomposition can be carried out by making the threefold rotation for momentum, spin, and curvature, as shown in Table~\ref{tab:silicon_curvature}. The SOI reads
\begin{align}
& H'_{(0, \pm k_0, 0)} \nonumber \\ 
& = \mp \frac{\hbar v_1}{R} s_y \sin 2\varphi 
\mp \frac{\hbar v_2}{R} s_x \sin^2\varphi 
\mp \frac{\hbar v_3}{R} s_x \cos^2\varphi.
\label{eq:silicon2}
\end{align}
For the $(0,0,\pm k_0)$-valley, in contrast, no curvature-induced term appears up to $1/R$. We have a nonzero term, $(\partial_x^2 + \partial_y^2) u_z \propto 1/R$, as an $A_1$ irrep, but this cannot be coupled to a same irrep, e.g., $k_z$, due to the PT symmetry.

\subsection{Tight-binding calculation}
\label{sec:calc-silicon}

\begin{figure}[tbp]
\centering
\includegraphics[width=\linewidth]{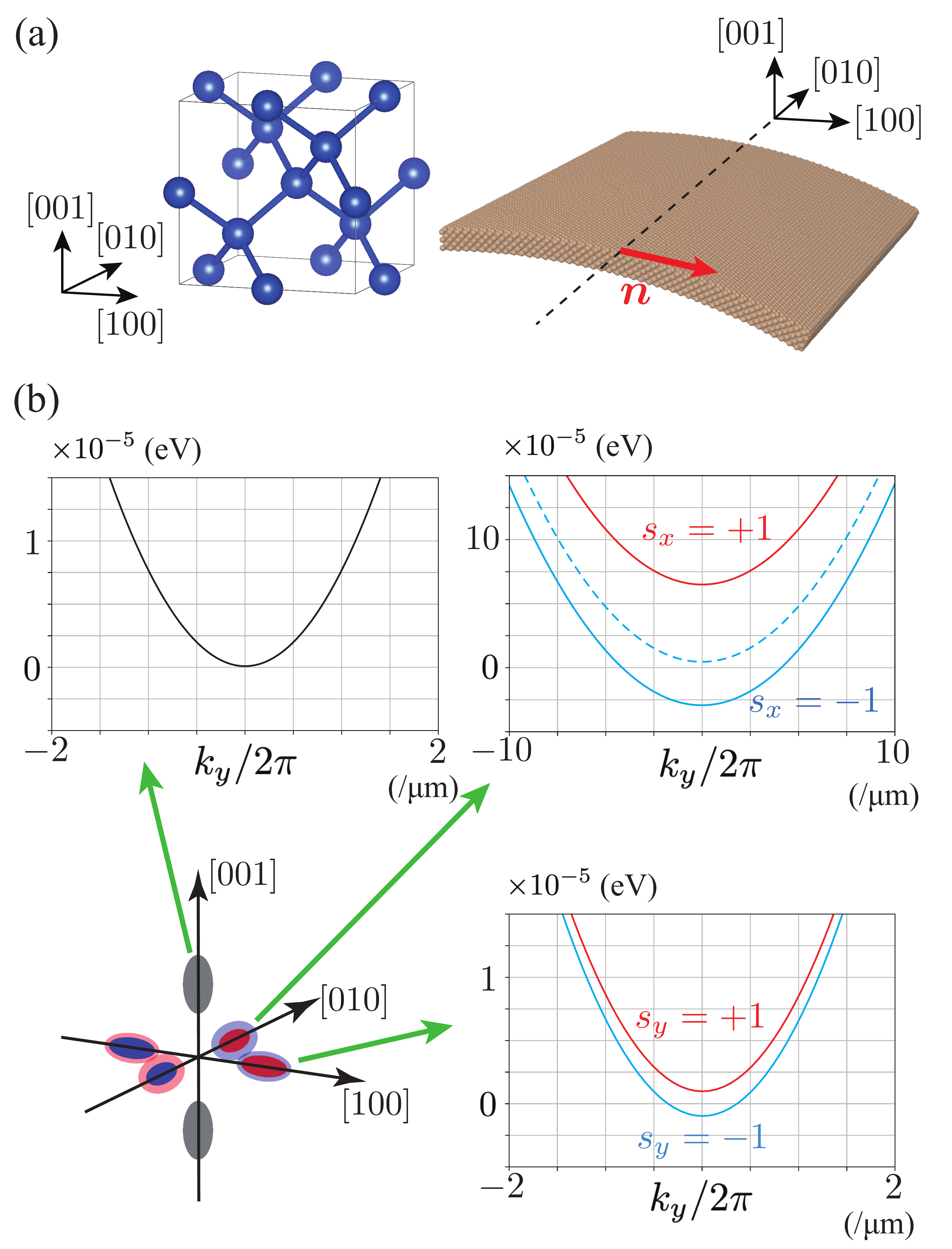}
\caption{(a) Crystal structure of silicon and schematic diagram of curved silicon whose bending direction is $\boldsymbol{n}=(1,0,0)$. (b) Schematic diagram of the valleys of silicon in the wavenumber space (lower left panel) and the conduction bands at the valleys of $[1 0 0]$ (lower right panel), $[0 1 0]$ (upper right panel) and $[0 0 1]$ (upper left panel) along the $k_y$ direction. The red and blue lines indicate band dispersions for different spin polarizations. The solid lines indicate the lowest subbands, while the dashed line indicates the next lowest subband. Here, the subbands are formed by imposing the boundary condition along the circumferential direction. The curvature radius is taken to be $R=50\,{\rm nm}$ and the energy is measured from the conduction band bottom for unbent silicon.}
\label{fig:calc_result}
\end{figure}
 
To estimate the magnitude of spin splitting, we calculate the band structure of curved silicon using the tight-binding model, taking into account the effect of the curvature up to order $1/R$. 
Let us consider curved silicon as schematically shown in Fig.~\ref{fig:calc_result}~(a), where the $[0 1 0]$ axis is kept unchanged, corresponding to $\varphi=0$. 
Figure~\ref{fig:calc_result}~(b) shows the calculated conduction bands near $(k_0,0,0)$, $(0,k_0,0)$, and $(0,0,k_0)$ with $R=50\,{\rm nm}$, where $R$ is the curvature radius of the neutral surface of a thin silicon.
At the valleys of $(k_0,0,0)$ and $(0,k_0,0)$, the electron spin is fully polarized along the $y$ and $x$ directions, respectively. The spin splitting of the conduction band is estimated to be $\sim 2\,\mu{\rm eV}$ at the valley of $(k_0,0,0)$ and $\sim 90 \,\mu{\rm eV}$ at the valley of $(0,k_0,0)$.
Note that the spin splitting at the valley of $(0,0,\pm k_0)$ is negligibly small. 
{These results are compared with the group-theoretic prediction for CNTs in Sec.~\ref{subsub:Hamiltonian}. }

Here, the spin $s_y$ and the momentum $k_c$ are not conserved, because they do not commute with the effective Hamiltonian near the valley.
Instead, the total angular momentum along the axis direction, $J_y = Rk_c + s_y/2$, becomes a good quantum number (see Eq.~(\ref{eq:spinorbit})), and therefore the cutting lines are labeled by $J_y$ {(see Appendix~\ref{app:calc})}.
Figure~\ref{fig:calc_result_S} shows the energy dispersion at the valley of $(0,k_0,0)$ as a continuous function of $J_y$. 
Note that $J_y$ is actually quantized by the boundary condition; $J_y=0$ ($J_y=\pm 1$) yields the lowest (next-lowest) cutting line, for example.

\begin{figure}[tbp]
\centering
\includegraphics[width=0.8\linewidth]{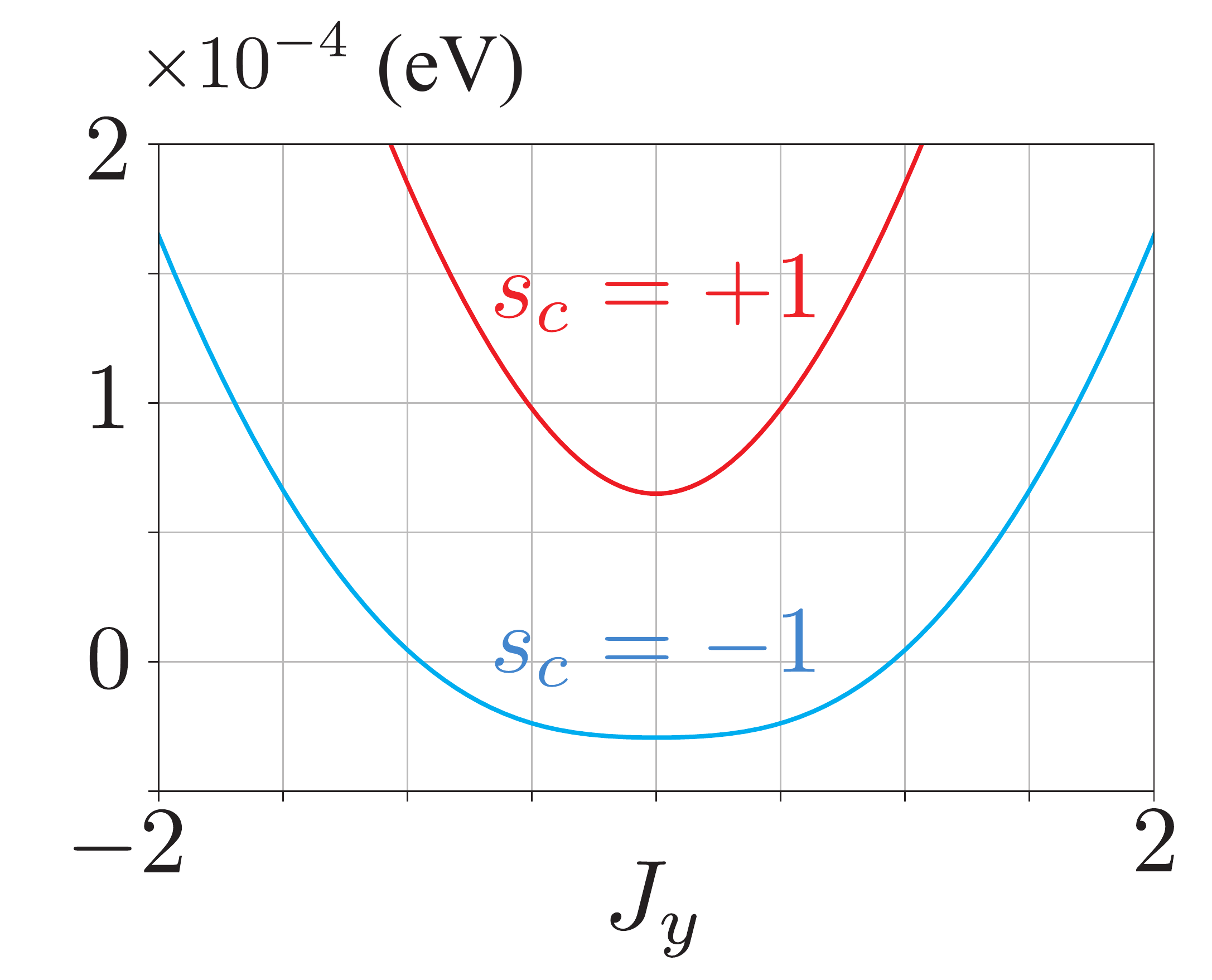}
\caption{Energy dispersion at the valley of $(0,k_0,0)$ for $\varphi=0$  as a function of angular momentum along the axis direction, $J_y$.}
\label{fig:calc_result_S}
\end{figure}
 
{The bending direction can be $\varphi \neq 0$.}
We also calculate the band structures when $\varphi=\pi/4$ by changing the directions of $x_t$, $x_c$, and $x_n$ with carefully treating the direction of a crystal structure. By comparing these numerical results with Eqs.~(\ref{eq:silicon}) and (\ref{eq:silicon2}), the values of $v_1,v_2,v_3$ can be estimated to be $\hbar v_1/R\simeq 30\,\mu{\rm eV}$, $\hbar v_2/R\simeq 1\,\mu{\rm eV}$, and $\hbar v_3/R\simeq 50\,\mu{\rm eV}$ for $R=50\,{\rm nm}$.

The spin splitting obtained here should be able to be detected with resonance microwave measurements~\cite{Sichau2019}.

\section{Thin-film quantization}
\label{sec:ThinFilmQuantization}

{We should note that the SOI induced in curved materials has been discussed in the literature in terms of the technique of thin-film quantization, which is a method to describe a particle confined in a one- or two-dimensional curved space embedded in three-dimensional space~\cite{daCosta1981,matsutani1992}.
In this method, the spatial constraint on a particle can be represented by the geometric potentials depending on the curvatures of the low-dimensional space.
Recently, this method has been applied to the chirality-induced spin selectivity~\cite{shitade2020geometric}.
In this section, we clarify that this method cannot be applied to curved materials in usual situations because the confinement potential has to be made unphysically large.}

{As an illustrative example, let us consider an effective Hamiltonian for CNTs by using the method given in Refs.~\cite{daCosta1981,matsutani1992}.}
For an electron confined in a curved thin film such as the CNT, one starts with a Dirac equation in 3+1 dimensional curved space-time with the geometric potentials of a cylindrical surface which is considered to be the CNT. 
Following Ref.~\cite{matsutani1992}, we use the cylindrical coordinates shown in Fig.~\ref{Fig2}. 
We start with the Dirac equation:
\begin{multline}
i\hbar \frac{\partial}{\partial t} \frac{\psi (x)}{\sqrt{1+x_n/R}}\\ 
= [ -i\hbar c \alpha^{\hat{i}}e_{\hat{i}}^{\ j} \partial_j + \beta mc^2 + V(x_n) ] \frac{\psi (x)}{\sqrt{1+x_n/R}},
\end{multline}
where $\alpha^{\hat{i}}$ is the alpha matrix, $V(x_n) = v_c x_n^{2}$ is the confinement potential to the nanotube with strength parameter $v_c$, and
$e_{\hat{i}}^{\ j}$ is the triad field, whose indices $\hat{i}$ and $j$ respectively label the Cartesian coordinates $(x,y,z)$ and cylindrical coordinates $(x_c,x_t,x_n)$:
\begin{align}
e_{\hat{i}}^{\ c} &= \frac{1}{\sqrt{1+x_n/R}} \left( -\sin \frac{x_c}{R} \delta_{\hat{i} z} + \cos \frac{x_c}{R}\delta_{\hat{i} x} \right), \\
e_{\hat{i}}^{\ t} &= \delta_{\hat{i} y}, \\
e_{\hat{i}}^{\ n} &= \cos \frac{x_c}{R}\delta_{\hat{i} z} + \sin \frac{x_c}{R} \delta_{\hat{i} x},
\end{align}
where $\delta_{{\hat{i}\hat{j}}}$ is Kronecker's delta.
{Note that $x_n=0$ when an electron is on the nanotube.}
By taking the confinement potential to be infinitely large, i.e., $v_c\rightarrow \infty$, we obtain the SOI modulated by the geometric potential in the non-relativistic limit as follows:
\begin{align}
H_{\text{so}}^{\text{tfq}} = \frac{\hbar s_t p_c}{2mR},
\end{align}
{where the superscript `tfq' indicates thin-film quantization.}

\begin{figure}[tb]
    \centering
    \includegraphics[width=50mm]{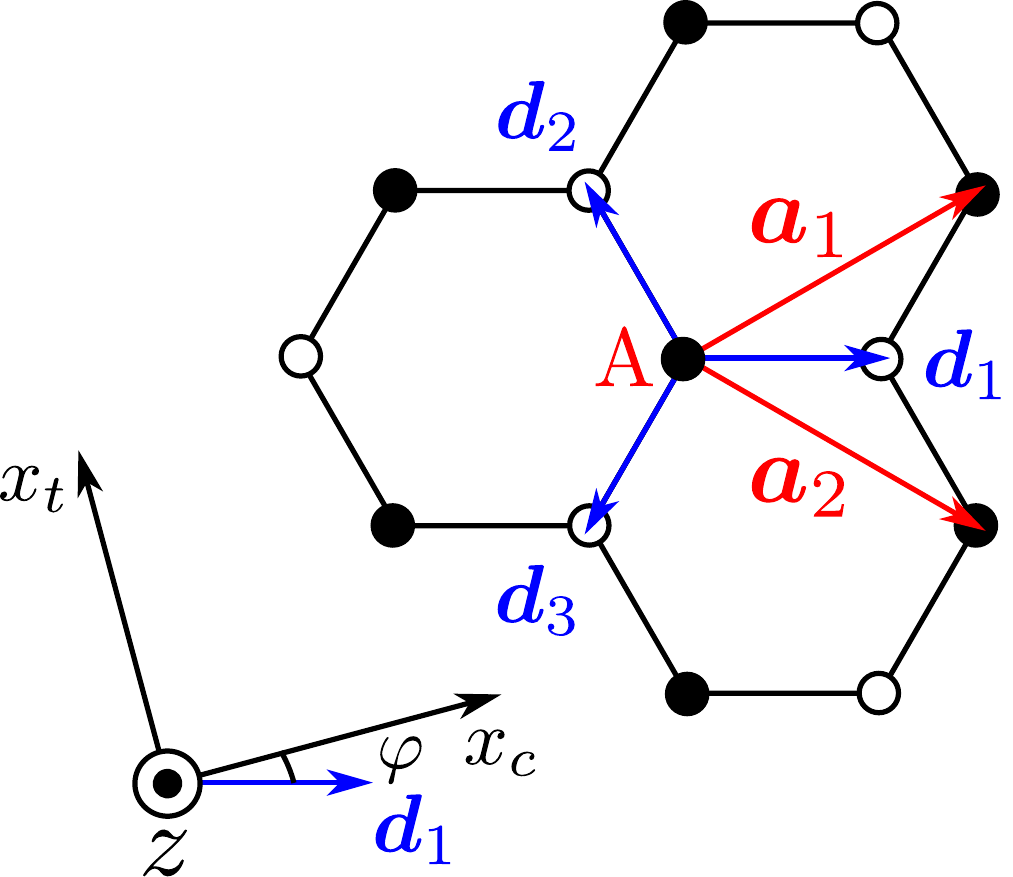}
    \caption{Crystalline structure of a CNT.
    $\bm a_1$ and $\bm a_2$ are primitive lattice vectors, and $\bm \Delta_j$ $(j=1,2,3)$ are vectors connecting the nearest-neighbor sites.
    $\theta$ is the angle between $\bm a_1$ and the $x_c$-axis.}
    \label{fig:tfq_graphene}
\end{figure}

Next, we estimate the SOI modulation derived by the thin-film quantization on a CNT (see Fig.~\ref{fig:tfq_graphene}).
The matrix elements of the SOI modulation between the nearest-neighbor sites on the CNT are given by
\begin{align}
\bra{\bm R_A} H^{\text{tfq}}_{\text{so}} \ket{\bm R_A+\bm d_j} = \frac{2\hbar v_F}{3} is_t \Delta k_{\text{so}}^{\text{tfq}} \cos \phi_j,
\end{align}
where $j=1,2,3$ are integers, $v_F\simeq 8.32 \times 10^5\,$m/s is the Fermi velocity, and $\phi_j = -\varphi + \frac{2\pi}{3}(j-1)$ is the angle between the displacement vector $\bm d_j$ and the $x_c$-axis.
$\Delta k^{\text{tfq}}_{\text{so}}$ characterizes the energy gap between the conduction and valence bands around $K$ and $K'$ points due to the SOI modulation: it is estimated as
\begin{align}
\Delta k_{\text{so}}^{\text{tfq}} = -\frac{v}{2v_FR} \simeq -0.82 \times \frac{1}{2R},
\label{eq:tfq_estimation}
\end{align}
where we have defined $v=\hbar \kappa/m$, and $2\kappa/3$ denotes the matrix elements of the momentum operator, given by~\cite{gruneis2003,gruneis_dthesis}
\begin{align}
\frac{2}{3}\kappa = \bra{\bm R_A} p_x \ket{\bm R_A+\bm d_j} \simeq 0.22\ \text{a.u.}
\end{align}
Our estimate of the curvature-induced SOI by the thin-film quantization is three orders of magnitude larger than the previous results, $\Delta k_{\text{so}}\simeq 5.3 \times 10^{-4} /2R$, and its sign is opposite to that of the SOI estimated by a first-principles calculation~\cite{Izumida2009-gx}.

This overestimation is caused by the constraint imposed by the infinitely large confinement potential $V(x_n)$, which is larger than the energy of the electron mass.
{This extreme condition is not fulfilled at all in the consideration of electronic states in materials.
Therefore, we conclude that the method of thin-film quantization should not be used for an estimate of the curvature-induced SOI as far as we start from the original Dirac equation for electrons.}

\section{Summary} 
\label{sec:Summary}

We proposed a group-theoretical method to describe the valley-dependent spin-orbit interaction induced in curved materials. This method systematically determines an effective Hamiltonian from the symmetries of the crystals, momenta, and spins.
The method succeeded in reproducing the effective Hamiltonian of CNTs with sublattice- and valley-dependent SOI induced by curvature. 
Furthermore, we derived an effective Hamiltonian for curved silicon and revealed that the curvature activates SOI in a valley-selective manner.
Combining this method with a tight-binding calculation, we demonstrated significant SOI splittings with nontrivial bending-direction dependences. 

{Our theory provides a concrete prospect on curvature-induced spin-orbit interactions far beyond a specific discussion for carbon nanotubes and silicon nanotubes.
In addition to materials discussed in this work, our theory can be utilized as a systematic method to produce valley-dependent SOI in various systems. For example, nanotubes produced by rolled-up atomic layer materials such as transition-metal dichalcogenides would be an interesting example~\cite{Musfeldt2020}. Our method can also be used to control the SOI in rolled-up semiconductor structures with different elastic properties~\cite{Prinz:2000aa,Zhang2017}.
Our theory is also applicable to curvature-induced SOI by using spatially inhomogeneous structures such as corrugations of atomic layers~\cite{avsar2020colloquium}.
Thus, our method will provide a general strategy for {\it curvature engineering} of valley-dependent SOI in nanomaterials.}
	
\begin{acknowledgments}
We acknowledge JSPS KAKENHI for providing Grants (No.~JP18H04282, No.~JP19K14637, No.~JP20K03831, No.~JP20H01863, No.~JP20K03835, No.~JP20K05258, No.~JP21K20356, {No.~JP21K03414, No.~JP24K06951}, No.~23H01839, and No.~24H00322) and the Sumitomo Foundation (190228). 
M.M. was supported by the National Natural Science Foundation of China (NSFC) under Grant No.~12374126 and by the Priority Program of Chinese Academy of Sciences under Grant No.~XDB28000000.T.S.~was supported by the Japan Society for the Promotion of Science through the Program for Leading Graduate Schools (MERIT).
\end{acknowledgments}

\appendix

\section{Details of Tight-Binding Calculation}
\label{app:calc}

{In this appendix, we formulate the hopping integral and the atomic SOI using the general coordinate transformation, Eq.~(\ref{eq:trans}).}

\subsection{Hopping integral}

For simplicity, we assume that the potential of an ion is the Coulomb potential, i.e., $V(x,y,z)=k/(x^2+y^2+z^2)^{1/2}$, where $k$ is the Coulomb constant.
Here, we take the Coulomb constant to be $k=Z\hbar c/137\epsilon_r$, where $c$ is the velocity of light, $Z$ is the effective ionic charge, and $\epsilon_r$ is the relative electrical permittivity.
We set $\epsilon_r=3$ for carbon nanotubes~\cite{Wang2012} and $\epsilon_r=11.7$ for silicon~\cite{Dunlap1953}, while we set $Z=4$ for both systems, of which results will be shown later.
If $R$ is much larger than $r$, the potential energy of an ion located at the origin can be written in the new coordinates $V(x',y',z')$ as
\begin{align}
&V(x',y',z')
=\frac{k}{\sqrt{R^2+r^2-2rR\cos\phi+y^2}} \nonumber
\\\nonumber&= \frac{k}{\sqrt{ x'^2+y'^2+z'^2}} \\
\nonumber &\hspace{5mm} \times \left(1-\frac{x'^2z'}{2R(x'^2+y'^2+z'^2)} +{\cal O}(1/R^2) \right) \\
&\equiv V_0(x',y',z')+\frac{1}{R}V_{\rm cur}(x',y',z') +{\cal O}(1/R^2).
\end{align}
Thus, in addition to the spherical Coulomb potential $V_0(x',y',z')$, we also obtain the curvature-induced anisotropic correction $V_{\rm cur}(x',y',z')/R$.

The present coordinate transformation also modifies the kinetic energy of electrons. 
The Laplace operator is expressed in cylindrical coordinates as
\begin{align}
\Delta &= \frac{1}{r}\partial_{r}\left(r\partial_r\right)+\frac{1}{r^2} \partial_\phi^2 +\partial_y^2.
\label{eq1}
\end{align}
Using $(x',y',z')=(R\phi, y,r-R)$, the spatial derivatives in the new coordinates are written in terms of $r$, $\phi$, and $y$ as
\begin{align}
\partial_{x'}=\frac{1}{R}\partial_\phi, \quad 
\partial_{y'}=\partial_y, \quad
\partial_{z'}=\partial_r.
\label{eq2}
\end{align}
Combining Eqs.~(\ref{eq1}) and (\ref{eq2}), the Laplace operator in the new coordinates becomes
\begin{align}
\Delta &\simeq \partial_{x'}^2+\partial_{y'}^2
+\partial_{z'}^2
+\frac{1}{R}\partial_{z'}-\frac{2z'}{R}\partial_{x'}^2 \nonumber 
\\&\equiv\Delta_0+\frac{1}{R}\Delta_{\rm cur},
\end{align}
where $\Delta_0 = \partial_{x'}^2+\partial_{y'}^2+\partial_{z'}^2$ is the Laplace operator in the new coordinates and $\Delta_{\rm cur}$ is a correction term induced by the coordinate transformation.

The wave functions are also modified by the coordinate transformation. For instance, the wave function of the $p_z$ orbital in the original frame is transformed as
\begin{align}
\nonumber
&\psi_{p_z}({\bm r})=\frac{r\cos\phi-R}{4\sqrt{2\pi a_z^5}}e^{-\sqrt{(r\cos\phi-R)^2+r^2\sin^2\phi+y^2}/2a_z} \\
\nonumber&\simeq \frac{1}{4\sqrt{2\pi a_z^5}} \left[z'+\frac{1}{R}\left(-\frac{x'^2}{2}-\frac{x'^2z'^2}{4a_z\sqrt{x'^2+y'^2+z'^2}}\right)\right]\\
\nonumber&\hspace{4cm}\times e^{-\sqrt{x'^2+y'^2+z'^2}/2a_z}
\\&\equiv \psi_{p_z}^{(0)}({\bm r}'
)+\frac{1}{R}\psi_{p_z}^{(1)}({\bm r}'
) ,
\end{align}
where $a_z=a_0/Z$, $a_0$ is the Bohr radius, $\psi_{p_z}^{(0)}({\bm r}') = \psi_{p_z}({\bm r}')$ is the original wavefunction, and $\psi_{p_z}^{(1)}({\bm r}')/R$ is a correction term induced by the coordinate transformation.

In our tight-binding calculation, we employ the non-orthogonal Slater-Koster two-center parameters~\cite{Min2006,papaconstantopoulos} for silicon and CNTs.
The correction of the hopping integral, which is proportional to $1/R$, is written as
\begin{widetext}
\begin{align}
\int d^3{\bm r}' &\left\{\psi^{(0)*}_n({\bm r}'+{\bm d}_i/2) \left(-\frac{\hbar^2}{2m}\Delta_{\rm cur}+ V_{\rm cur}({\bm r}'+{\bm d}_i/2)+V_{\rm cur}({\bm r}'-{\bm d}_i/2)\right)\psi^{(0)}_m({\bm r}'-{\bm d}_i/2)\right. \nonumber \\
&\left.+\psi^{(1)*}_n({\bm r}'+{\bm d}_i/2) \left(-\frac{\hbar^2}{2m}\Delta_0+ V_0({\bm r}'+{\bm d}_i/2)+V_0({\bm r}'-{\bm d}_i/2)\right)\psi^{(0)}_m({\bm r}'-{\bm d}_i/2)\right. \nonumber \\
&\left.+\psi^{(0)*}_n({\bm r}'+{\bm d}_i/2) \left(-\frac{\hbar^2}{2m}\Delta_0+ V_0({\bm r}'+{\bm d}_i/2)+V_0({\bm r}'-{\bm d}_i/2)\right)\psi^{(1)}_m({\bm r}'-{\bm d}_i/2)\right.\nonumber\\
&\left.+
z'\psi^{(0)*}_n({\bm r}'+{\bm d}_i/2) \left(-\frac{\hbar^2}{2m}\Delta_0+ V_0({\bm r}'+{\bm d}_i/2)+V_0({\bm r}'-{\bm d}_i/2)\right)\psi^{(0)}_m({\bm r}'-{\bm d}_i/2)\right\}, \label{integ}
\end{align}
\end{widetext}
where the displacement vectors between two atomic orbitals are denoted with ${\bm d}_i$ ($i=1,\cdots,M$). 
{We note that the volume integral in the old coordinates is rewritten with the new coordinates as
\begin{align}
\int dx \, dy \, dz \, (\cdots)
= \int dx'\, dy'\, dz' \, \frac{r}{R}(\cdots).
\end{align}
The last term of Eq.~(\ref{integ}) is due to the correction in the Jacobian, $r/R$.}

\subsection{Spin-orbit interaction (SOI)}

{In our tight-binding calculation, we take the atomic SOIs into account through one-site terms.}
The atomic SOI is defined in cylindrical coordinates, which is $(x_c,x_t,x_n)$ in Fig.~\ref{Fig2} and is given as $\lambda {\bm s}\cdot{\bm l}$, where ${\bm s}=(s_c,s_t,s_n)$ is the spin operator and ${\bm l}=(l_c,l_t,l_n)$ is the orbital angular momentum operator in cylindrical coordinates.
Since the spin operator is related to the spin of an itinerant electron, ${\bm s}$ has to be rewritten in the laboratory frame $(x,y,z)$ (see Fig.~\ref{Fig2}). 
When the axis of the nanotube is taken to be in the $y$ ($=y'$) direction, the spin operator in cylindrical coordinates, ${\bm s}$, can be rewritten as~\cite{Izumida2009-gx}
\begin{align}
s_c &=(\tilde{s}_+\mu_- - \tilde{s}_-\mu_+)/i, \\
s_t&=\tilde{s}_y, \\
s_n&=\tilde{s}_+\mu_-+\tilde{s}_-\mu_+,
\end{align}
where $\tilde{{\bm s}}=(\tilde{s}_x,\tilde{s}_y,\tilde{s}_z)$ is the spin operator in the laboratory frame, $\tilde{s}_{\pm} = \tilde{s}_z \pm i \tilde{s}_x$ is the spin ladder operator, and $\mu_\pm$ is the operator which changes the momentum of the circumference direction $k_c$ by $\pm\delta k_c=\pm 1/R$. Using these relations, the atomic SOI takes the form of 
\begin{align}
\label{eq:spinorbit}
\lambda {{\bm s}} \cdot {\bm l} = \lambda(\tilde{s}_yl_y+ \tilde{s}_+\mu_-l_-+\tilde{s}_-\mu_+l_+).
\end{align}
This spin-orbit interaction hybridizes a up-spin state with wavenumber $(k_c,k_t,k_n)$ with a down-spin state with wavenumber $(k_c\pm \delta k_c,k_t,k_n)$.
This transfer between $k_c$ and $\tilde{s}_y$ is ascribable to the conservation of the total angular momentum in the $y$ direction in the laboratory frame, as we will discuss soon.
For example, let us consider a band calculation of silicon using eight atomic orbitals, i.e., four atomic orbitals ($3s$, $3p_x$, $3p_y$, and $3p_z$) per sublattice.
Here, we need to consider a $16\times 16$ matrix for the Hamiltonian $\mathcal{H}({\bm k})$ by taking the spin degree of freedom into account, in which the eight up-spin orbitals with wavenumber $(k_c,k_t,k_n)$ couple to the eight downspin orbitals with shifted wavenumber $(k_c+\delta k_c,k_t,k_n)$.
Since $s_c$ and $s_n$ change the (quantized) 
momentum in the circumference direction, $k_c$, it is no longer a good quantum number. Instead, the total angular momentum in the direction of the $y$-axis, $J_y=Rk_c+s_y/2$, is conserved, due to the axial symmetry in the laboratory frame. 
We can obtain the band structure by numerically diagonalizing the Hamiltonian $\mathcal{H}({\bm k})$ for fixed ${\bm k}$.

For nanotube structures, we should also take into account the boundary condition for the circumference direction. In the absence of the SOI, $k_c$ is discretized in order to satisfy the boundary condition. Therefore, the original Brillouin zone for an unbent material is quantized into line segments~\cite{Saito-1998}, called cutting lines. A cutting line is nothing but a quasi-one-dimensional subband, labeled by $k_c$. In the presence of the SOI, $J_y$ should be used instead of $k_c$ to specify a subband~\cite{Izumida2009-gx}, as $k_c$ is no longer a good quantum number.
	
\bibliography{./reference}

\begin{thebibliography}{56}
\bibitem [{foo()}]{footnote}%
  \BibitemOpen
  \href@noop {} {}\bibinfo {howpublished} {Unlike $K_6$ irrep ($\pi$ electrons) for graphene, $\Delta_1$ irrep is non-degenerate, and therefore orbital degrees of freedom, such as $\boldsymbol{\sigma}$ for graphene, do not appear in this case.}\BibitemShut {Stop}%
\end{thebibliography}
	
\clearpage

\end{document}